# Accelerating micromagnetic and atomistic simulations using multiple GPUs

Serban Lepadatu[1]

[1]*Jeremiah Horrocks Institute for Mathematics, Physics and Astronomy, University of Central Lancashire, Preston PR1 2HE, U.K.*

**Abstract**

It is shown micromagnetic and atomistic spin dynamics simulations can use multiple GPUs to reduce computation time, but also to allow for a larger simulation size than is possible on a single GPU. Whilst interactions which depend on neighbouring spins, such as exchange interactions, may be implemented efficiently by transferring data between GPUs using halo regions, or direct memory accesses, implementing the long-range demagnetizing interaction is the main difficulty in achieving good performance scaling, where the data transfer rate between GPUs is a significant bottleneck. A multi-GPU convolution algorithm is developed here, which relies on single-GPU FFTs executed in parallel. It is shown that even for micromagnetic simulations where the demagnetizing interaction computation time dominates, good performance scaling may be achieved, with speedup factors up to 1.8, 2.5, and 3.1, for 2, 3, and 4 GPUs respectively. The code developed here can be used for any number of GPUs in parallel, with performance scaling strongly dependent on inter-GPU data transfer rate and connection topology. This is further improved in micromagnetic simulations which include a spin transport solver, obtaining speedup factors up to 1.96, 2.8, and 3.7, for 2, 3, and 4 GPUs respectively. The best case scenario is obtained for atomistic simulations, where the demagnetizing interaction is implemented with spin-averaged cells. Using a single workstation with 4 GPUs, it is shown atomistic spin dynamics simulations with up to 1 billion spins, and atomistic Monte Carlo simulations with up to 2 billion spins are possible, with a near-ideal performance scaling.

*SLepadatu@uclan.ac.uk

# I. Introduction

Micromagnetic and atomistic spin dynamics simulations have become essential for analysis of many experimental results on magnetic materials and samples, as well as design and modelling tools for advanced spintronics devices and applications. Examples include neuromorphic computing [1], skyrmionic neural networks [2], magnetic data storage and processing [3,4], heat-assisted magnetic recording [5], spin wave computing and magnonics [6,7], and nano-oscillators [8]. OOMMF [9] was one of the first open-source micromagnetics software, implementing finite-difference micromagnetics for computations on central processing units (CPU). Graphical processing units (GPU) allowed a significant increase in performance due to massive parallelisation [10,11], following which Mumax1 [12] was the first GPU-based open-source software implementing finite-difference micromagnetics. Currently there are a number of software, including Mumax3 [13], Fidimag [14], but also Python-based magnum.np [15] and Ubermag meta-package [16]. For atomistic spin dynamics modelling, open-source software include Vampire [17] and Spirit [18], with Fidimag also allowing atomistic modelling. BORIS [19] is another finite-difference open-source software, allowing both advanced micromagnetic modelling with multi-physics capabilities, as well as atomistic and multiscale modelling on CPUs and GPUs, with simulation control done through Python scripts, either via network sockets or through the embedded Python interpreter. All publically available software currently run on single GPUs, which places a limit on the complexity of problems which can be simulated, both in terms of size and performance. The step from CPU to GPU computation was important, but equally important is the step from single-GPU to multi-GPU computation. Workstations and data servers with multiple GPUs are now widely available, and it is imperative for modelling software to unlock the full potential of modern hardware. It should be noted that multi-GPU methods have been introduced in other fields, notably in computational electromagnetics where Poisson's equation is solved using fast Fourier transforms (FFT) with good scalability up to 16 GPUs [20], and multi-GPU parallelization of 3D FFTs was also discussed [21].

The step from single-GPU to multi-GPU computation however is not trivial. Whilst interactions involving differential operators are relatively easy to implement, the biggest difficulty is implementing the long-range demagnetizing interaction, particularly since this can account for up to 90% of the total computation time. The difficulty arises since every computational cell requires the contribution from all other computational cells be included, which is particularly difficult for a multi-GPU implementation where bandwidth is an important bottleneck to performance. This problem is solved here by introduction of a multi-GPU convolution algorithm, allowing the demagnetizing field to be computed efficiently across any number of GPUs. The multi-GPU capabilities discussed here, including micromagnetics modelling, atomistic spin dynamics and Monte Carlo algorithm, as well multi-physics capabilities, including a heat equation solver, thermo-elastodynamics solver, and spin transport drift-diffusion solver, have been implemented in an upgraded BORIS codebase available as open source [22].

## II. Multi-GPU Convolution

The demagnetizing field, $\mathbf{H}_d$, is given by a convolution sum at cell $i$ as:

$$\mathbf{H}_{d,i} = -\sum_{j} \mathbf{N}(i-j) \mathbf{M}_j \tag{1}$$

$\mathbf{N}$ is the demagnetizing tensor, computed using the formulas in Ref. [23], and $\mathbf{M}$ is the magnetization, with the sum running over all points in the magnetic mesh. Whilst Equation (1) is straightforward to evaluate directly, both on a single node as well as on multiple nodes – e.g. see Ref. [17] for a multi-CPU implementation – its complexity scales as $O(N^2)$, where N is the number of computational cells, making a naïve evaluation impractical for large mesh sizes when using GPUs. Instead, the convolution theorem may be used so that $\mathbf{N}$ and $\mathbf{M}$ are first transformed using a FFT, followed by point-by-point multiplication in the complex transform space, and finally obtaining $\mathbf{H}_d$ using an inverse FFT (IFFT). The complexity of this approach scales as $O(N\log N)$, and is a standard evaluation method for single-GPU and CPU finite-difference implementations. There are a number of ways such a convolution could be implemented across multiple GPUs. Three methods have been implemented and tested here, including a method based on the multi-layered convolution algorithm previously introduced for multi-mesh single-GPU computation [24], where data is transferred between GPUs in the complex transform space before point-by-point multiplication, as well as a simplistic method where the $\mathbf{M}$ data is fully exchanged between GPUs. The best performing method, in all cases, is depicted in Figure 1. Here, the computational workload is equally split between the different GPUs, and the code developed in this work is applicable for any number of GPUs from 2 upwards – Figure 1 exemplifies the algorithm for 3 GPUs, although it has been tested on 2, 3, and 4 GPUs. The general case of the algorithm is given in Appendix A in pseudo-code form.

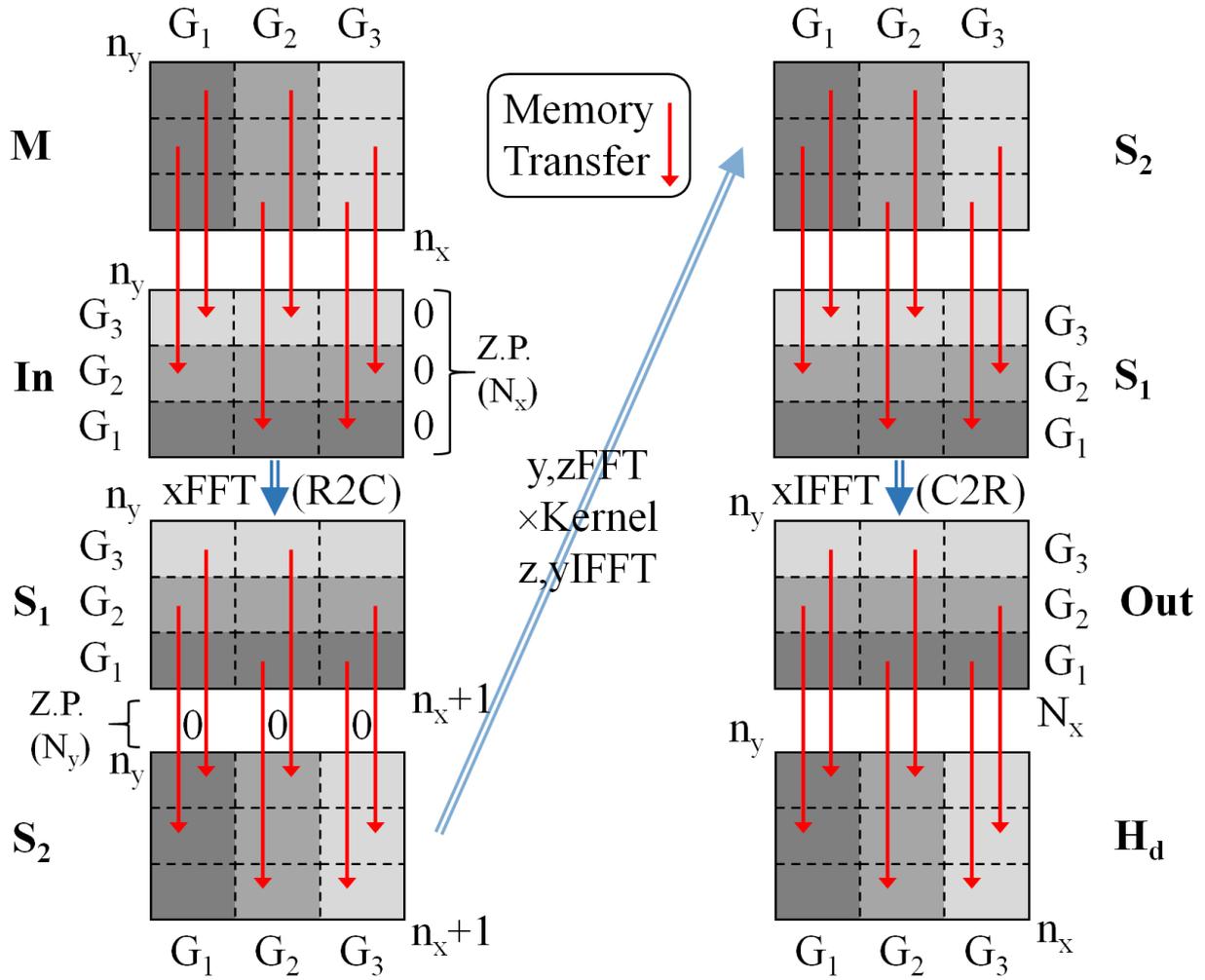

**Figure 1** – Multi-GPU convolution algorithm outline, exemplified for 3 GPUs. **M** and $\mathbf{H}_d$ are the magnetization and demagnetizing field respectively, with mesh dimensions $n_x \times n_y \times n_z$. Rows and columns of same grayscale shading denote contiguous memory spaces on same GPU, as labelled by $G_1$, $G_2$, and $G_3$ respectively. Prior to the x FFT step memory is transferred between GPUs into zero-padded (to $N_x = 2n_x$) input spaces, **In**, so that the FFTs are of length $N_x$, but each GPU performs a third of the total number of FFTs. The x FFT output is stored in complex-valued spaces $\mathbf{S}_1$, with x dimension $n_x+1$. Prior to the y and z FFT steps, memory is transferred between GPUs into zero-padded (to $N_y = 2n_y$, $N_z = 2n_z$) complex spaces, $\mathbf{S}_2$, so that full length FFTs are performed, but each GPU does a third of the work. Following point-by-point convolution kernel multiplications, the IFFT pipeline is similar to the forward FFT pipeline, but performed in reverse. Memory transfers are denoted using red arrows, where the floating point precision can be halved before transfers in order to reduce the bandwidth used.

The **M** space of dimensions $n_x \times n_y \times n_z$ is partitioned equally between the different GPUs along the x dimension (approximately, since if the number of GPUs does not integer-divide $n_x$ the last GPU is required to have a slightly different size along x). The forward FFTs are performed first along x, then y, and finally z directions, to full length as in the single-GPU implementation. However, since there are multiple FFTs along each dimension, the workload can be split equally between the different GPUs, which for the x FFT step is achieved by partitioning the y dimension equally between the different GPUs (again approximately, since the number of GPUs may not integer-divide $n_y$) – for the y and z FFT steps the x dimension partitioning is used. Thus, before the x FFT step it is necessary to transfer data between the different GPUs, as shown in Figure 1, between the **M** and **In** spaces – the **In** spaces are now partitioned along the y dimension between the different GPUs, and additionally contain zero padding from $n_x + 1$ to $N_x = 2n_x$, which is required due to circular convolution for a finite sum when periodic boundary conditions (PBC) [25] are not used. Since the input data are real the x FFTs are of real-to-complex (R2C) type, with output stored in the complex space $S_1$ of dimension $n_x + 1$ along x. Before the y and z FFT steps may be performed, it is now necessary to transfer data between GPUs, from $S_1$ to $S_2$ spaces, so that the x dimension partitioning is used instead. The $S_2$ space is also zero padded along the y and z dimensions, up to $N_y = 2n_y$ and $N_z = 2n_z$ respectively. Following forward FFTs, point-by-point multiplication with the transformed demagnetizing tensor – termed convolution kernel – is performed. The kernel is computed in the initialization stage, as for the single-GPU implementation, however it is partitioned between the different GPUs so that only the required data are stored on each GPU respectively. The IFFT pipeline is similar to the forward FFT pipeline, but performed in reverse. The main difference is after the x IFFT step which, being of complex to real (C2R) type, results in real output data in the **Out** space, of size $N_x$ along the x dimension, and partitioned between the different GPUs along the y dimension. Memory transfers are finally performed from **Out** to $H_d$, also truncating the output since points stored in **Out** from $n_x + 1$ to $N_x$ are not required, resulting in demagnetizing field values stored in the correct place for each GPU with x dimension partitioning.

Since the workload is split equally between the GPUs, good performance scaling could be expected. There are 3 main factors which affect this. Latency is introduced when launching computational routines on GPUs. The total latency increases with number of GPUs used, however this is typically a bigger issue with smaller problem sizes. We also require an overhead compared to the single-GPU case, as data must be copied into contiguous memory before transfer, then again copied into computational spaces after transfer. The more important factor is due to data transfers between GPUs, which is the most significant bottleneck for almost all problem sizes. The algorithm of Figure 1 requires that each GPU sends/receives ~$18N(N_G-1)/N_G^2$ floating point numbers each iteration, which scales as $O(N/N_G)$, where $N_G$ is the number of GPUs. Thus, for a given mesh size, if the GPUs are all interconnected using a point-to-point connection topology, e.g. NVSwitch, latency and overhead aside, the algorithm should theoretically increase in efficiency by increasing the number of GPUs. At the other extreme, for the simpler bus connection topology, the more relevant metric is the total floating point numbers transferred over the bus, which is $\sim 18N(N_G-1)/N_G$ and thus the total floating point numbers transferred is limited to less than $18N$ in all cases.

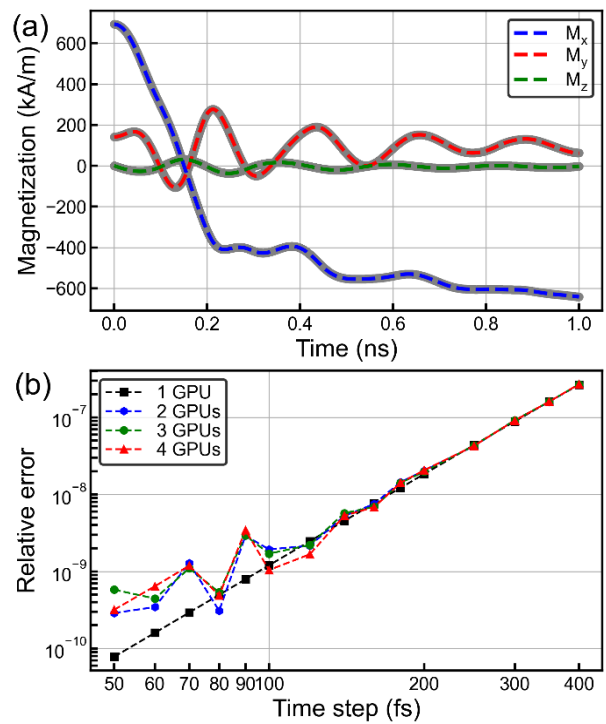

**Figure 2** – Error scaling with time step for the RK4 evaluation method, relative to a reference solution computed with 10 fs time step on a single GPU in double precision. (a) Micromagnetic switching problem used to compute the error scaling, showing the reference solution using dashed lines, and solution computed with 400 fs time step and 4 GPUs in single precision with half precision transfers, using solid thick lines. (b) Relative error scaling computed for 1, 2, 3, and 4 GPUs, where the data transfer between GPUs is done using halved precision.

One method of reducing the total data transferred between GPUs is to use mixed precision. This requires the floating point precision be halved before transfer, e.g. for computations done in double precision, single precision transfers are used, and for computations done in single precision, half precision transfers are used, which results in a loss of precision. It is important to normalize data before transfer in order to avoid reaching the exponent limit, particularly for half precision. **M** data are normalized to $M_s$, the saturation magnetization, whilst complex-valued FFT data are normalized to N×$M_S$. Figure 2 shows the relative error computed as a function of time step using the Runge-Kutta 4$^{th}$ order evaluation method (RK4) [26], for a micromagnetic switching problem with Zeeman, exchange, and demagnetizing interactions. The mesh size is 2.4 µm × 1.2 µm × 10 nm, discretized with a 5 nm cellsize, where $M_S$ = 800 kA/m. The magnetization is initialized in an *S* state by saturation in a 1 MA/m field along the [1, 1, 1] direction, which is gradually reduced to 0. Following this, a switching field is applied as **H** = (-20, 1, 0) kA/m, and the average magnetizing is recorded every time step for 1 ns. Figure 2(a) shows the computed switching event, where a reference solution is computed in double precision with a 10 fs time step on a single GPU, shown using dashed lines. At the other extreme, the solution obtained using 4 GPUs with 400 fs time step in single precision, and with half precision transfers, is shown using the solid thick lines. The solutions are virtually identical, with $R^2$ > 0.999. The solutions can be quantitatively compared by computing a mean relative error, defined as $\varepsilon = \sum_{i=1}^{n} \left\| \mathbf{M}(t_i) - \tilde{\mathbf{M}}(t_i) \right\| / nM_S$, where $\tilde{\mathbf{M}}(t_i)$ are the points in the reference dataset, and $\mathbf{M}(t_i)$ are the computed *n* points over the 1 ns interval. This is shown in Figure 2(b) for 1, 2, 3, and 4 GPUs, with computations done in double precision and transfers done in single precision. It can be seen the error introduced is negligible, particularly for time steps greater than 100 fs. Similar tests were done with computations in single precision, and transfers in half precision. In this case the lower precision limits the accuracy to which the relative error can be computed, which is ~$10^{-4}$ for all time steps, both for single-GPU and multi-GPU computations, again showing no significant error is introduced by use of properly normalized halved precision transfers. Finally, the code developed here also includes PBCs, which requires a modified convolution kernel as in the single-GPU implementation [25], and no zero padding along the PBC direction.

A further test is discussed, this time using a perpendicular magnetic tunnel junction (MTJ) where the effect of spin torques is simulated. The structure used consists of a 6 nm thick antiferromagnetic (AFM) layer, a 3 nm thick permanent magnetic layer (PL), a 1 nm thick tunnel barrier (I), and a 7 nm thick free magnetic layer (FL). This is shown in Figure 3(a), where an MTJ dot with 100 nm diameter is simulated, discretized using a 1 nm cellsize – the RK4 method with 20 fs time-step was used. The layers have magnetic parameters $M_S$ = 636 kA/m, exchange stiffness 10 pJ/m, uniaxial perpendicular anisotropy 318 kJ/m$^3$, and a Gilbert damping constant 0.01. Additionally, the layers are coupled using surface exchange coupling, namely exchange bias coupling between PL and AFM with 1 mJ/m$^2$ coupling constant, and a weaker surface exchange coupling between FL and PL with 0.1 mJ/m$^2$ coupling constant – the same interactions implemented in Ref. [27] are used here, with further details also given in the Supplementary Material.

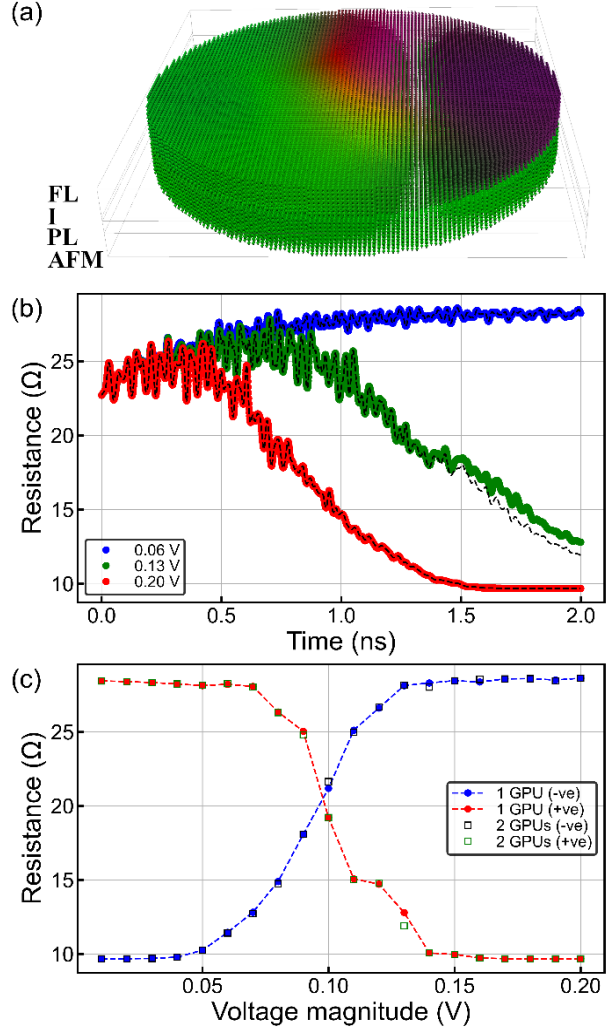

**Figure 3** – Switching pulsed voltage calculation in a perpendicular MTJ, comparing results obtained using 1 GPU in single precision, with 2 GPUs in single precision and half precision transfers. (a) MTJ dot with 100 nm diameter, and layer thicknesses AFM (6 nm) / PL (3 nm) / I (1 nm) / FL (7 nm). (b) Individual switching events for different voltages, with solid lines showing single-GPU computations, and dashed lines showing computations with 2 GPUs. (c) Computed resistance after a voltage pulse for both positive (anti-parallel starting state) and negative (parallel starting state) voltages.

A voltage is applied between the top and bottom faces in Figure 3(a) (the bottom face is the ground electrode), such that for a voltage of 0.1 V, a current density of $J_c = 10^{11}$ A/m$^2$ flows into the ground electrode. This results in a spin torque on the FL, given by $\mathbf{T}_{ST} = (\mu_B/e)(J_c/d_{FL})\eta \mathbf{m} \times (\mathbf{m} \times \mathbf{p})$, where for simplicity the field-like spin torque is set to zero. Here $d_{FL}$ is the thickness of FL, **m** is the magnetization direction in FL, **p** is the magnetization direction in PL, and $\eta = 0.25$ is a spin torque efficiency factor. The tunnelling magnetoresistance (TMR) of the MTJ is calculated using Slonczewski's formula [28] with an MTJ RA product of $0.15 \times 10^{-12}$ $\Omega m^2$ and 200% TMR. A 2 ns voltage pulse is applied – examples of dynamic resistance calculations are shown in Figure 3(b), where computations using 1 GPU in single precision are compared with computations using 2 GPUs with half precision transfers. A relatively large MTJ structure was chosen so that during magnetization switching a multi-domain structure emerges, as exemplified in Figure 3(a), which allows for a more stringent test of the multi-GPU convolution algorithm with half precision transfers. The computed resistance values after a 2 ns voltage pulse are shown in Figure 3(c). Positive voltages switch the MTJ from the anti-parallel to the parallel configuration, and negative voltages switch it from the parallel to the anti-parallel configuration. Due to the short voltage pulse the switching threshold is not abrupt, leading to a transition region as shown in Figure 3(c) – this was purposely chosen to allow for a more detailed test. The same transition regions are obtained for computations with 1 and 2 GPUs. For most simulations the dynamics are nearly identical for 1 and 2 GPUs, however for a few simulations discrepancies can arise, as seen in Figure 3(b) for the 0.13 V voltage pulse. Since magnetization switching occurs along an axis with rotational symmetry, it is possible that such divergences can occur at local maxima points in the energy landscape under the influence of reduced precision and hence different floating point errors. In general, it is recommended that the suitability of half precision transfers for the particular problem being studied is verified through comparison with fully single-precision calculations, before a full set of simulations is performed, and this should only be enabled for bandwidth-restricted setups.

# III. Micromagnetic Simulations

Single-site interactions, such as anisotropy – e.g. Equations (2) and (3) show the uniaxial and cubic anisotropy fields respectively – are trivial to parallelize across multiple GPUs, and the same x dimension partitioning from Figure 1 is used.

$$\mathbf{H}_{an} = \frac{2K_1}{\mu_0 M_S}(\mathbf{m}.\mathbf{e}_A)\mathbf{e}_A + \frac{4K_2}{\mu_0 M_S}[1-(\mathbf{m}.\mathbf{e}_A)^2](\mathbf{m}.\mathbf{e}_A)\mathbf{e}_A \qquad (2)$$

$$\begin{aligned}\mathbf{H}_{an} = &-\frac{2K_1}{\mu_0 M_S}[\mathbf{e}_1\alpha(\beta^2+\gamma^2)+\mathbf{e}_2\beta(\alpha^2+\gamma^2)+\mathbf{e}_3\gamma(\alpha^2+\beta^2)] \\ &-\frac{2K_2}{\mu_0 M_S}[\mathbf{e}_1\alpha\beta^2\gamma^2+\mathbf{e}_2\alpha^2\beta\gamma^2+\mathbf{e}_3\alpha^2\beta^2\gamma]\end{aligned} \qquad (3)$$

In Equations (2) and (3) $\mathbf{e}_A$ is the uniaxial symmetry axis, $\mathbf{m} = \mathbf{M} / M_S$, $K_1$ and $K_2$ are anisotropy constants, and $\alpha = \mathbf{m}.\mathbf{e}_1$, $\beta = \mathbf{m}.\mathbf{e}_2$, and $\gamma = \mathbf{m}.\mathbf{e}_3$, where $\mathbf{e}_3 = \mathbf{e}_1 \times \mathbf{e}_2$, with $\mathbf{e}_1$, $\mathbf{e}_2$, $\mathbf{e}_3$ being the cubic symmetry axes. The iteration time for the cubic anisotropy interaction is shown in Figure 4(a), as a function of number of computational cells, for 1, 2, 3, and 4 GPUs. The computational platform used was a single workstation running Ubuntu 20.04, with 4 Nvidia A5000 24GB GPUs, connected via PCIe 4.0, each with an x16 slot (32 GB/s total bandwidth) – the CPU used, namely Ryzen Threadripper PRO 3955WX, provides a total of 128 PCIe 4.0 control lanes. The GPUs are also connected directly in pairs using NVLink bridges, which provides a separate higher 56.248 GB/s bandwidth (a comparison of performance with and without additional NVLink bridges is given in the Supplementary Material). Computations are done here in single precision, as is common practice with micromagnetic modelling (e.g. Ref. [13]). For a large enough problem size the iteration time decreases by increasing the number of GPUs used, reaching a near ideal speedup efficiency of ~0.97 in all cases. Efficiency is defined as (S – 1) / ($N_G$ – 1), where S is the speedup, defined as the iteration time for single-GPU computation, divided by the iteration time for computation with $N_G$ GPUs. This can reach a maximum of 1 (ideal efficiency), and can be negative for speedup values less than 1. Plots of speedup efficiency for the interactions in Figure 4 are given in the Supplementary Material. For small problem sizes it is observed the iteration time reaches a lower limit, which becomes more pronounced by increasing the number of GPUs. This is due to GPU latencies as discussed in the previous section.

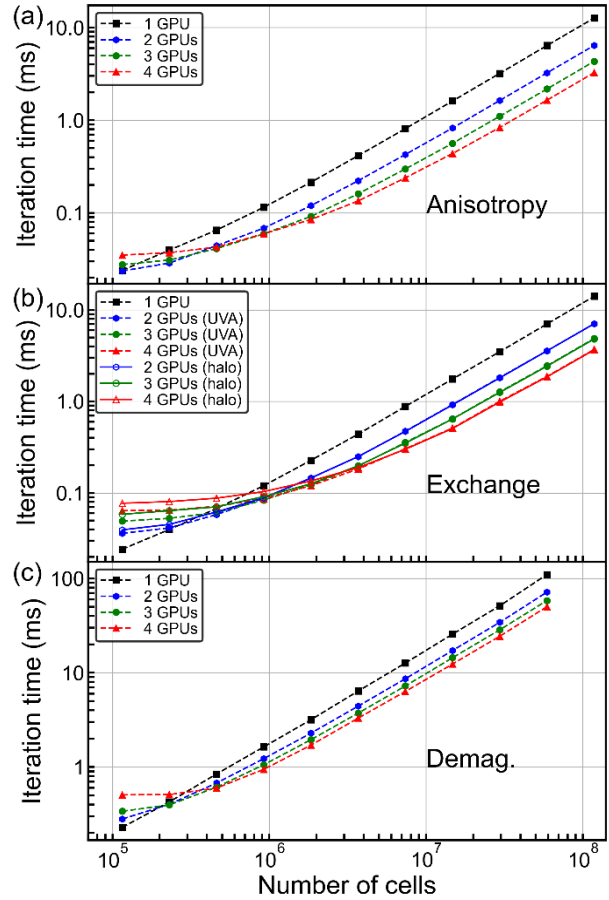

**Figure 4** – Iteration time of various interactions as a function of number of computational cells, for 1, 2, 3, and 4 Nvidia A5000 GPUs, for (a) cubic anisotropy, (b) direct exchange, and (c) demagnetizing interaction. For the exchange interaction the halo transfer and UVA methods of computing differential operators at GPU partition boundaries are compared.

Another important case is that of local interactions involving neighbours, or in general equations involving differential operators. This is the case for the direct exchange interaction, which is given in Equation (4) where $A$ is the exchange stiffness:

$$\mathbf{H}_{ex} = \frac{2A}{\mu_0 M_S} \nabla^2 \mathbf{m} \tag{4}$$

There are 2 methods used to evaluate differential operators at GPU partition boundaries. One method is to define halo regions either side of the partition, and transfer them between relevant

GPUs before a computation. A further method involves direct memory accesses, which for Nvidia GPUs may be done using the unified virtual addressing (UVA) architecture. Computational routines are launched on a target GPU, however through UVA it is possible to address memory on other GPUs within the same computational routine. Thus, with UVA, halo regions are not required. The iteration times for the direct exchange interaction are shown in Figure 4(b). As with the anisotropy interaction, for small problem sizes the iteration time reaches a lower limit, which again is largely due to GPU latencies. For large problem sizes the efficiency increases (see Supplementary Material), reaching values of 0.97, 0.95, and 0.94 for 2, 3, and 4 GPUs respectively, both for the halo and UVA methods. Whilst the UVA and halo methods provide the same performance for large problem sizes, the UVA method is significantly better than the halo method for small problem sizes, as see in Figure 4(b). This is due to additional latencies for the halo method, where memory must be copied to contiguous memory spaces before transfer, then copied back into computational spaces after transfer. There is however a case when the halo method outperforms the UVA method, namely for 3 or more GPUs when purely a bus connection topology is used. In Figure 4(b) the GPUs are connected via the PCIe bus, but also by NVLink bridges in pairs. When a GPU is required to read data from 2 other GPUs via PCIe, with UVA a drop in performance was observed as shown in the Supplementary Material. Whilst memory transfers are done directly between GPUs over the PCIe bus (peer-to-peer i.e. not via CPU host memory), this is still a serial point-to-point connection, which requires CPU control to switch contexts. It is likely this additional overhead is the cause for poor performance when using UVA over PCIe.

Finally, the iteration times for the demagnetizing interaction, computed using the multi-GPU convolution algorithm, are shown in Figure 4(c). As for the exchange interaction GPU latencies and memory transfers limit the performance for small problem sizes, however due to the much larger cost of memory transfers, the maximum efficiencies are also lower, namely reaching 0.54, 0.44, and 0.40 for 2, 3, and 4 GPUs respectively. Nevertheless, the multi-GPU convolution algorithm does provide a significant computational speedup (1.54, 1.89, and 2.19 for 2, 3, and 4 GPUs respectively – lower speedup factors of 1.44, 1.56, and 1.70 respectively are obtained when using full precision transfers, and for completeness this is also discussed in the Supplementary Material), and efficiencies are expected to improve with greater bandwidth in more advanced computational platforms, or when using a point-to-point connection topology, as discussed in the previous section. Moreover, the demagnetizing interaction is very rarely required on its own, and from a practical perspective more relevant speedup factors and efficiencies are obtained when considering all interactions in typical micromagnetic problems. The interaction fields are added to obtain a total effective field, $\mathbf{H}_{eff}$, which is included in the Landau-Lifshitz-Gilbert (LLG) equation, where $\alpha$ is the Gilbert damping and $\gamma$ is the gyromagnetic ratio, as:

$$\frac{\partial \mathbf{m}}{\partial t} = -\gamma \mathbf{m} \times \mathbf{H}_{eff} + \alpha \mathbf{m} \times \frac{\partial \mathbf{m}}{\partial t} \tag{5}$$

For typical micromagnetic problems we have $\mathbf{H}_{eff} = \mathbf{H}_{Zee} + \mathbf{H}_{an} + \mathbf{H}_{ex} + \mathbf{H}_{d}$, where $\mathbf{H}_{Zee}$ is the Zeeman term (applied external field). The LLG equation evaluation, as well as the Zeeman term, with multiple GPUs is also trivial, as with the anisotropy interaction, since $\mathbf{m}$ and $\mathbf{H}_{eff}$ are already partitioned along the x dimension. The LLG and Zeeman term iteration times added together are similar to the cubic anisotropy interaction, thus looking at Figure 4, the demagnetizing interaction evaluation time accounts for ~80 to 90% of the total computation time. The LLG equation is typically evaluated using a higher order explicit evaluation method, e.g. RK4, or for adaptive time stepping the Runge-Kutta-Fehlberg $5^{th}$ order method with $6^{th}$ order error estimation (RKF56) [29]. There is a method of reducing the overall demagnetizing interaction computation time, without decrease in solution accuracy, which is based on polynomial extrapolation of the demagnetizing field at sub-steps of the higher order evaluation method, using previously computed values. This was discussed in a previous work [30], and in particular the RK4 method requires a quartic polynomial, whilst the RKF56 method requires a quintic polynomial. This method is particularly beneficial for multi-GPU implementation, since by decreasing the time spent on the less efficient demagnetizing interaction computation (decrease by a factor between 2 to 2.5 is possible), the overall efficiency can be increased.

To test this, the same magnetization switching problem of Figure 2 is used, but also including the cubic anisotropy interaction, and with the LLG equation evaluated using the RKF56 method. The speedup factors for 2, 3, and 4 GPUs are calculated as a function of number of computational cells, which is achieved by increasing the in-plane mesh dimensions in factors of 2. The results are shown in Figure 5. The maximum speedup factors reached are 1.82, 2.47, and 3.14, for 2, 3, and 4 GPUs respectively (for full precision transfers the speedup factors are 1.79, 2.31, and 2.80 respectively).

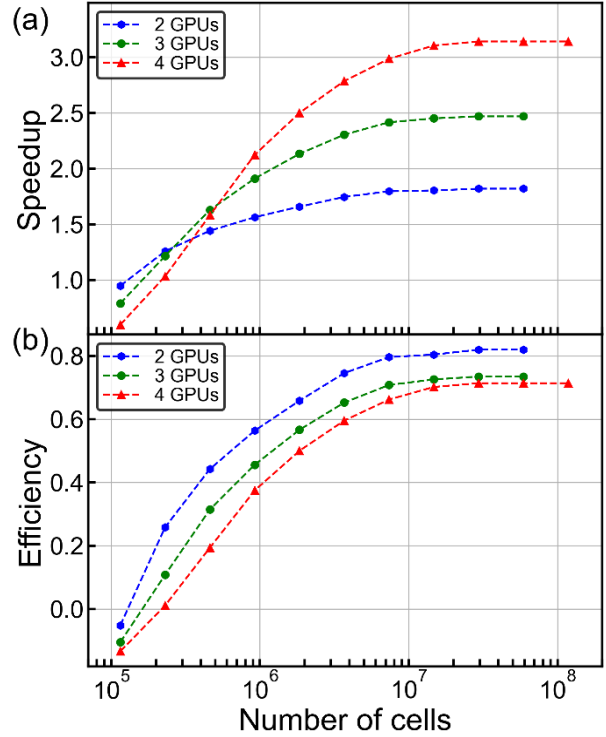

**Figure 5** – Computational speedup for 2, 3, and 4 GPUs over single-GPU computation, for a micromagnetic problem with Zeeman, direct exchange, cubic anisotropy, and demagnetizing interactions, as a function of number of computational cells, showing (a) speedup, and (b) speedup efficiency. The LLG equation is solved using the RKF56 method and quintic polynomial extrapolation for the demagnetizing interaction.

From Figure 5(b) it is observed the drop in efficiency from 3 to 4 GPUs is less than the drop between 2 and 3 GPUs. This is consistent with the increase in total floating point numbers transferred by increasing the number of GPUs, which is given by $18N(N_G-1)/N_G$, i.e. $9N$, $12N$, and $13.5N$ for 2, 3, and 4 GPUs respectively. Thus, whilst the efficiency drops, this is expected to converge to a constant efficiency for large problem sizes by increasing the number of GPUs, since the total floating point numbers transferred is limited below $18N$, although further testing is required to confirm this. The largest problem which can be executed on a single GPU has ~30 million cells, whilst for 4 GPUs the largest possible problem has ~120 million cells (speedup factors for problems with more than 30 million cells are calculated by extrapolating the single-GPU computation times). Use of multiple GPUs is particularly beneficial for large problems (defined here as having more than 10 million cells), with large speedup factors possible, but also enabling larger problem sizes which cannot be executed on a single GPU. Even for medium-sized problems (defined here as having between 1 and 10 million cells), use of multiple GPUs is beneficial, and also for small problems (less than 1 million cells) use of 2

GPUs is advantageous. This is particularly important from the user point of view, since for modern workstations configurations with 2 GPUs can be easily achieved. As computational platforms improve, the code developed in this work should perform with greater efficiencies as inter-GPU bandwidth increases.

A simple model of speedup parallelization may be constructed. Using the number of floating point values transferred each iteration, and denoting the number of bytes per floating point number as $B$ (4 for single precision, 2 for half precision), and the bandwidth as $R$, the time required each iteration to transfer data between GPUs connected using a serial bus is given by $t_R(N_G) = 18(N/N_G)(N_G - 1)(B/R)$. Then, if $t_I$ is the time spent evaluating interactions which can be parallelized efficiently (Zeeman, anisotropy, exchange, LLG evaluation), and $t_D$ is the time spent evaluating the demagnetizing interaction, the total time for each iteration is $t_T(N_G) = (t_I + t_D(r_D + r_{Dextrap}))/\eta N_G + t_R r_D$. Here $\eta$ is an efficiency factor ($<1.0$), and $r_D$, $r_{Dextrap}$ are demagnetizing evaluation time reduction factors introduced when using the polynomial extrapolation method. Thus for RKF56 $r_D = 1/8$ since the method contains 8 sub-steps per iteration, and $r_{Dextrap} = 0.1$ is an additional factor introduced due to polynomial extrapolation computation; for RK4 we have $r_D = 1/4$ and $r_{Dextrap} = 0.05$. The efficiency $\eta$ can be estimated by fitting the measured interaction times for the results up to 4 GPUs. Thus for a problem with N = 58,982,400 computational cells, measured interaction times are $t_I$ = 26.8 ms, $t_D$ = 122.3 ms, and $\eta$ = 0.98 is obtained. This model reproduces the speedup factors measured in Figure 5, which we can then use to extrapolate a speedup factor of less than 5 for 8 GPUs. This shows the benefit of using an increasing number of GPUs is limited when bandwidth is restrictive. Useful predictions may be obtained for other simulation platforms. For example PCIe 5 allows $R$ = 64 GB/s, with a modelled speedup factor of ~7.5 for 16 GPUs, whilst for the upcoming PCIe 6 platform with $R$ = 128 GB/s a speedup factor of ~10 for 16 GPUs is predicted. Finally, the NVSwitch platform allows full interconnections between all GPUs, which means $t_R$ can also be parallelized, and is now given by $t_R(N_G) = 18(N/N_G^2)(N_G - 1)(B/R)$. The NVSwitch 3$^{rd}$ generation platform allows for 8 GPUs to be interconnected, each with $R$ = 900 GB/s bandwidth (7.2 TB/s aggregate bandwidth), however multiple NVSwitch boards can also be interconnected. Using the fully single precision convolution algorithm ($B$ = 4), and no polynomial extrapolation for the demagnetizing field evaluation ($r_D$ = 1 and $r_{Dextrap}$ = 0), a near-ideal speedup factor of over 15 is predicted for 16 GPUs (plots of modelled speedup factors are given in the Supplementary Material).

# IV. Atomistic Simulations

For atomistic modelling we also have the same interactions discussed above, although defined differently, and a multi-GPU implementation is similarly achieved. The uniaxial anisotropy field is given as:

$$\mathbf{H}_{an} = \frac{2K_1}{\mu_0 \mu_B \mu_S}(\hat{\mathbf{S}}.\mathbf{e}_A)\mathbf{e}_A + \frac{4K_2}{\mu_0 \mu_B \mu_S}[1-(\hat{\mathbf{S}}.\mathbf{e}_A)^2](\hat{\mathbf{S}}.\mathbf{e}_A)\mathbf{e}_A \tag{6}$$

$\mathbf{S} = \mu_S \hat{\mathbf{S}}$ is the atomistic spin, with direction $\hat{\mathbf{S}}$ and magnitude $\mu_S$ (units of Bohr magneton, $\mu_B$). The direct exchange interaction field at a given spin $i$ is obtained by considering nearest neighbours, and is given as:

$$\mathbf{H}_{ex,i} = \frac{J}{\mu_0 \mu_B \mu_S} \sum_{j \in N} \hat{\mathbf{S}}_j \tag{7}$$

Here $J$ is the exchange interaction energy, and the sum runs over the nearest neighbours of spin $i$. Whilst this does not involve a differential operator as with the micromagnetic exchange interaction, Equation (4), the multi-GPU implementation is similar, also requiring a halo or UVA method. Whilst a simple cubic lattice was used here, with atomistic modelling farther neighbours may be included with different exchange energies, and for different crystal structures, and the implementation could be easily extended either by use of UVA, or by use of larger halo regions – this will be explored in a future work. We also have a long-range interaction between spins, termed the dipole-dipole interaction, given as:

$$\mathbf{H}_{d-d,i} = \frac{\mu_B}{4\pi} \sum_{i \neq j} \frac{3(\mathbf{S}_j.\hat{\mathbf{r}}_{ij})\hat{\mathbf{r}}_{ij} - \mathbf{S}_j}{r_{ij}^3} \tag{8}$$

Here $\mathbf{r}_{ij} = r_{ij}\hat{\mathbf{r}}_{ij}$ is the distance vector from spin $i$ to spin $j$, with the sum running over all spins in the atomistic lattice. The dipole-dipole interaction was also implemented using a convolution sum since Equation (8) can be re-arranged into the form of Equation (1). Thus, the multi-GPU performance scaling of atomistic models which include the dipole-dipole interaction is nearly identical to that obtained for micromagnetic modelling. However, in many cases it is not necessary to include the full dipole-dipole interaction, and a coarser spin-averaged cell may be used to compute the demagnetizing field of Equation (1). The magnetization is obtained as the total spin in a coarser cell which contains an integer number of lattice constants in each dimension, divided by the cell volume. This is given as:

$$\mathbf{M}_i = \frac{1}{h^3} \sum_j \mathbf{S}_j = \frac{1}{a^3 N_S} \sum_j \mathbf{S}_j \qquad (9)$$

Here $\mathbf{M}_i$ is the spin-averaged cell magnetization with cubic cellsize $h$, and $a$ is the atomistic lattice constant such that the spin-averaged cell contains $N_S$ spins. Since a coarser cell is used to evaluate the demagnetizing field, its share of the total computation time can be reduced by over an order of magnitude compared to a full micromagnetic model. The computed demagnetizing field is applied equally to all spins in the spin-averaged cell, and as for micromagnetic modelling the total effective field is a sum of the separate interaction fields. This is included in the stochastic LLG (sLLG) equation, allowing non-zero temperature modelling, as:

$$\frac{\partial \mathbf{S}}{\partial t} = -\gamma \mathbf{S} \times (\mathbf{H} + \mathbf{H}_{th}) + \frac{\alpha}{\mu_S} \mathbf{S} \times \frac{\partial \mathbf{S}}{\partial t} \qquad (10)$$

The thermal field, $\mathbf{H}_{th}$, follows a Gaussian distribution with zero mean and standard deviation given by $H_\sigma = \sqrt{2\alpha k_B T_e / \gamma \mu_0 \mu_B \mu_S \Delta t}$. Here $T_e$ is the electron bath temperature and $\Delta t$ is the evaluation method time step.

An example large-scale atomistic spin dynamics problem is shown in Figure 6. Here, the effect of an ultrafast laser pulse on a 16 nm thick FePt film ($\mu_S$ = 3.6 $\mu_B$, $K_1$ = 6.4×10$^{-23}$ J with easy axis perpendicular to the film, $J$ = 6.78×10$^{-21}$ J resulting in a Curie temperature of 710 K; and $\alpha$ = 0.1) is computed with in-plane dimensions of ~1.4 μm. Using a 0.4 Å lattice constant results in a problem size with ~0.5 billion spins. The sLLG equation is solved using the RK4 method with 8 fs time step. The demagnetizing field is computed in spin-averaged cells with 1.6 nm cellsize.

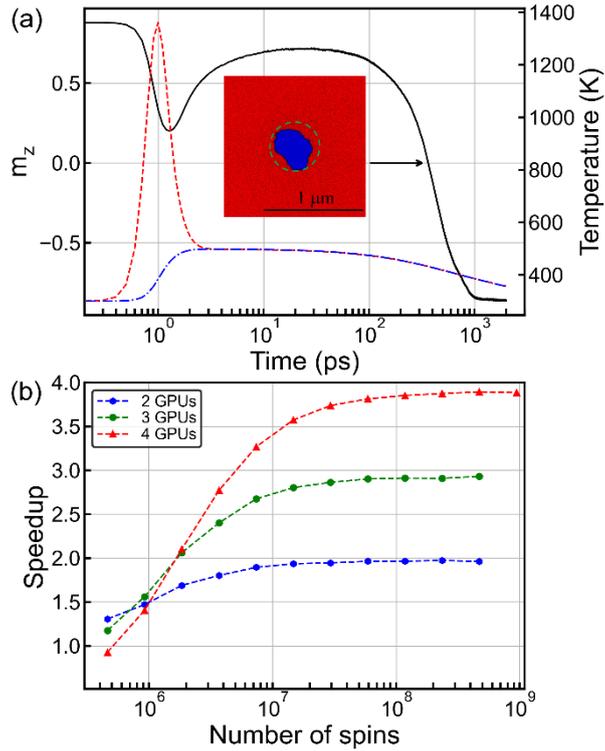

**Figure 6** – Ultrafast demagnetization and magnetization reversal under a uniform external field, computed using atomistic spin dynamics coupled to the 2-temperature heat equation, with Zeeman, direct exchange, uniaxial anisotropy, and demagnetizing interactions, in a 16 nm thick FePt film with 0.4 Å lattice constant. The sLLG equation is solved using the RK4 method and quartic polynomial extrapolation for the demagnetizing interaction. (a) $m_z$ component of normalized magnetization as a function of time for a problem with ~0.5 billion spins computed using 4 GPUs, in response to an ultrafast laser pulse with Gaussian temporal and spatial profiles, with size indicated in the inset with the dashed green circle. The electron bath and lattice temperatures are shown using the dashed red line and dash-dot blue line respectively. The inset shows the spins z components, with red positive and blue negative, at 400 ps. (b) Computational speedup for 2, 3, and 4 GPUs over single-GPU computation as a function of number of atomistic spins.

It is known that an ultrafast laser pulse can cause rapid demagnetization, followed by a subsequent magnetization recovery [31]. On the ultrafast timescale (femtosecond timescale) it is known the conduction and lattice electrons respond on very different timescales, which is modelled through different specific heat capacities [32], and a two-temperature model is used, given as:

$$C_e\rho\frac{\partial T_e(\mathbf{r},t)}{\partial t} = K\nabla^2 T_e(\mathbf{r},t) - G_e[T_e(\mathbf{r},t) - T_l(\mathbf{r},t)] + S(\mathbf{r},t)$$
$$C_l\rho\frac{\partial T_l(\mathbf{r},t)}{\partial t} = G_e[T_e(\mathbf{r},t) - T_l(\mathbf{r},t)]$$
(11)

Here $C_e$ and $C_l$ are the electron and lattice specific heat capacities, $G_e$ is the electron-lattice coupling constant, typically of the order $10^{18}$ W/m³K, $\rho$ is the mass density, and $K$ is the thermal conductivity. The electron and lattice temperatures are $T_e$ and $T_l$ respectively. Equation (11) is evaluated using the forward time centred-space scheme with 2 fs time step and 1.6 nm cellsize. We have $C_e$ = 40 J/kgK, $C_l$ = 430 J/kgK, $K$ = 46.4 W/mK, and $\rho$ = 8740 kg/m³. Since Equation (11) only involves differential operators and local contributions, its multi-GPU implementation is straightforward, with temperature memory spaces also using an x dimension partitioning between the different GPUs. The ultrafast laser pulse is introduced through the heat source $S$ in Equation (11), which has spatial and temporal Gaussian profiles as:

$$S = S_0 \exp\left(-\frac{2r^2}{r_0^2}\right)\exp\left(-\frac{8(t-2\tau)^2}{\tau^2}\right)$$
(12)

Here $S_0$ = 2.5×10²¹ W/m³ is the power density, $2r_0$ = 0.5 μm is the spot size, and $\tau$ = 400 fs is the pulse width. The effect of the laser pulse is seen in Figure 6(a): the electron bath temperature increases rapidly on a fs timescale, with temperatures far exceeding the Curie temperature. After the laser pulse the electron bath temperature decreases rapidly as it equilibrates with the slower changing lattice temperature. The temperature then decreases slowly on a much longer timescale as heat is lost to the ambient room temperature. The initial increase in temperature results in a rapid demagnetization, as seen in Figure 6(a), where the average $m_z$ component of normalized magnetization (computed from the atomistic spins in the laser spot region) is plotted. After the laser pulse the magnetization is recovered as the temperature decreases. Here a uniform external field of 1 MA/m is applied into the plane, which results in magnetization switching on a longer timescale as seen in Figure 6(a). Note, this is similar to the mechanism used for heat-assisted magnetic recording (HAMR) [5], although for magnetic recording granular films are used, rather than a continuous film, and the external field is also localized. Here, in the continuous FePt film magnetization switching is obtained as a reverse domain is nucleated in the laser spot region, which then slowly grows under the action of the external magnetic field. The speedup factors are plotted in Figure 6(b) as a function of number of spins used (varied by changing the in-plane FePt film simulated dimensions). The maximum speedup efficiencies obtained are ~0.95 to 0.97 for 2, 3, and 4 GPUs. The maximum problem size which could be simulated on a single GPU had ~0.25 billion spins, whilst for 4 GPUs the maximum problem size was close to 1 billion spins. From a performance point of view, the simulation of Figure 6(a) was completed in ~24 h with the 4 Nvidia A5000 GPUs, and looking at the efficiency factors it is clear such large-scale problems could benefit from additional GPUs. Using the speedup parallelization model introduced in the previous section, with $t_R(N_G) = 18(N/N_G)(N_G - 1)(B/R)$ ($N_S = 64$ here due to use of spin-averaged cells for the demagnetizing interaction) a near-ideal speedup factor of ~7.5 is predicted for 8 GPUs. This potentially allows micromagnetic sized problems [33,34] to be simulated using atomistic resolution.

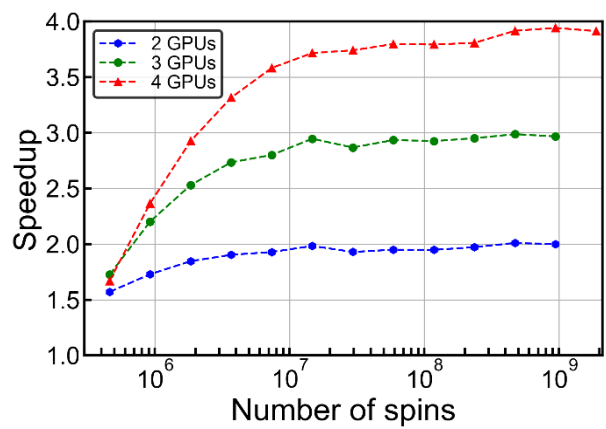

**Figure 7** – Computational speedup for 2, 3, and 4 GPUs over single-GPU computation, for an atomistic Monte Carlo problem with Zeeman, direct exchange, and cubic anisotropy interactions, as a function of number of computational cells.

As a final note, multi-GPU implementation of Monte-Carlo simulations is straightforward, with the relevant interactions already discussed in this section. The multi-GPU implementation uses the red-black checkerboard decomposition scheme [35], previously used to implement the Monte Carlo algorithm on a single GPU [36], and such a scheme has already been implemented for multi-GPU acceleration of the 2D Ising model [37]. The same method is used to implement the micromagnetic Monte Carlo method [38] on multiple GPUs, which additionally includes the demagnetizing interaction. The speedup factors obtained for the atomistic Monte Carlo method are shown in Figure 7, where maximum efficiency factors between 0.94 to 0.98 are obtained for 2, 3, and 4 GPUs. The maximum problem on a single GPU was ~0.5 billion spins, whilst 4 GPUs allow problem sizes up to 2 billion spins.

# V. Multi-Physics Simulations

Another area where use of multiple GPUs is beneficial, is for simulations which include a relatively computationally expensive multi-physics model. The heat equation was used in the previous section for atomistic modelling, however iterating Equation (11) is cheap, comparable in computational cost to the exchange interaction. Another possibility is the use of a thermo-elastodynamics solver, recently introduced into BORIS [39], which is more computationally expensive, but since it only involves single-site and nearest-neighbour interactions is parallelized using multiple GPUs efficiently. A more computationally expensive model is the spin transport drift-diffusion model [40,41], particularly since this typically requires cellsize values smaller than the magnetic cellsize, and with the successive over-relaxation (SOR) method multiple spin transport solver iterations are required for each LLG equation iteration. In the drift-diffusion model we have charge and spin currents, $\mathbf{J}_C$ and $\mathbf{J}_S$ respectively:

$$\mathbf{J}_C = \sigma \mathbf{E} + \theta_{SHA} D_e \frac{e}{\mu_B} \nabla \times \mathbf{S}$$

$$\mathbf{J}_S = -\frac{\mu_B}{e} P \sigma \mathbf{E} \otimes \mathbf{m} - D_e \nabla \mathbf{S} + \theta_{SHA} \frac{\mu_B}{e} \boldsymbol{\varepsilon} \sigma \mathbf{E}$$

(13)

Here $\sigma$ is the electrical conductivity, $\mathbf{E} = -\nabla V$ is the electric field with $V$ the electric potential, $\theta_{SHA}$ is the spin Hall angle in a heavy metal (HM) which leads to the spin Hall effect (SHE) [42], $D_e$ is the electron diffusion constant, $P$ is the current spin polarization, and $\boldsymbol{\varepsilon}$ is the rank 3 unit antisymmetric tensor. The spin accumulation, $\mathbf{S}$, follows the equation of motion, where $\lambda_{sf}$ is the spin-flip length, $\lambda_J$ is the exchange rotation length, and $\lambda_\varphi$ is the spin dephasing length:

$$\frac{\partial \mathbf{S}}{\partial t} = -\nabla \cdot \mathbf{J}_S - D_e \left( \frac{\mathbf{S}}{\lambda_{sf}^2} + \frac{\mathbf{S} \times \mathbf{m}}{\lambda_J^2} + \frac{\mathbf{m} \times (\mathbf{S} \times \mathbf{m})}{\lambda_\varphi^2} \right)$$

(14)

On the timescale of magnetization processes we are interested in the steady-state solution of Equation (14), thus by setting this to zero and by using Equation (13) with the condition of zero charge current divergence, we obtain 2 Poisson equations in $V$ and $\mathbf{S}$ (e.g. see Ref. [41]) which are solved here using the SOR method to a set convergence factor ($10^{-5}$ normalized convergence error for *V* and **S** equations is used). Using the computed spin accumulation we can calculate bulk and interfacial spin torques which can be included in the LLG equation as additional torque terms. In particular, the bulk spin torque in a ferromagnetic (FM) layer is given as:

$$\mathbf{T}_S = -\frac{D_e}{\lambda_J^2}\mathbf{m}\times\mathbf{S} - \frac{D_e}{\lambda_\varphi^2}\mathbf{m}\times(\mathbf{m}\times\mathbf{S}) \qquad (15)$$

The interfacial spin torque is obtained using the complex spin mixing conductance, $G^{\uparrow\downarrow}$, as:

$$\mathbf{T}_S = \frac{g\mu_B}{eh_F}\left[\text{Re}\{G^{\uparrow\downarrow}\}\mathbf{m}\times(\mathbf{m}\times\Delta\mathbf{V}_S) + \text{Im}\{G^{\uparrow\downarrow}\}\mathbf{m}\times\Delta\mathbf{V}_S\right] \qquad (16)$$

Here $h_F$ is the discretization cellsize in the ferromagnetic layer of a FM/HM interface, and $\Delta\mathbf{V}_S$ is the spin chemical potential change at the interface computed using boundary conditions for the charge and spin currents (see Refs. [41,43] for further details). The self-consistently computed total spin torques in Equations (15) and (16) contain 3 important different types of spin torques, among others. One of them is the bulk spin-transfer torque (STT) [44], which results from Equation (15). Two other spin torques are of interfacial type, arising from Equation (16). The first is the spin-orbit torque (SOT) which arises due to the SHE. In a HM layer the SHE gives rise to a pure spin current in a direction orthogonal to a charge current. Thus, for a FM/HM bilayer, a spin current flows from the HM layer into the FM layer, carrying spin angular momentum which results in a spin torque on the magnetization in the FM layer. When magnetization gradients are present a local spin accumulation is generated in the FM layer, which results in an imbalance in spins between the FM and HM layers. This results in a diffusive vertical spin current, this time flowing from the FM to the HM layer. As spin angular momentum is carried away from the FM layer, due to conservation of total angular momentum a spin torque arises on the FM layer, termed the interfacial STT (ISTT), which has been shown experimentally by studying the motion of skyrmions [45]. Experimental studies on current-induced Néel skyrmion motion – disk-like topological objects in magnetization textures [46] which can be stabilized using the interfacial Dzyaloshinskii-Moriya interaction (DMI) [47,48] at FM/HM interfaces – typically involve large sample sizes and multilayers [49-51], which can make associated computational studies expensive and could benefit from multi-GPU acceleration. An example is shown in Figure 8, where a multilayer structure consisting of Pt(2.8)/[Co(0.8)/Ir(0.4)/Pt(0.6)]$_{\times 8}$/Pt(2.2) – thicknesses in nm – is used, similar to the experimental samples in Ref. [45]. With an experimental track width of 2 μm, an 8 μm track length requires ~0.4 billion cells for the spin transport solver (2 nm in-plane cellsize, but a 0.2 nm cellsize is required along the z direction to resolve spin accumulation gradients), which requires 3 or more A5000 GPUs.

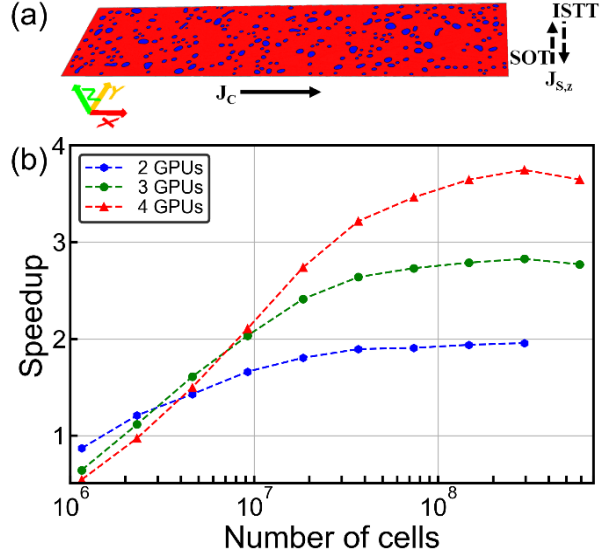

**Figure 8** – Skyrmion motion computed using a micromagnetic model with Zeeman, interfacial DMI, direct exchange, uniaxial anisotropy, surface roughness, and demagnetizing interactions, coupled to the drift-diffusion spin transport solver in a Pt(2.8)/[Co(0.8)/Ir(0.4)/Pt(0.6)]$_{\times 8}$/Pt(2.2) multilayer. (a) The multilayer track with 2 μm width and 8 μm length, with charge current along the track. A number of spin torques are present, namely SOT due to a z-direction spin current originating in the HM layers, ISTT due to a spin current originating at magnetization gradients in Co, as well as bulk STT in Co. (b) Computational speedup for 2, 3, and 4 GPUs over single-GPU computation as a function of number of computational cells.

The simulated track in Figure 8(a) contains a skyrmion collection, stabilized using the DMI with effective field shown below, where *D* is the DMI constant:

$$\mathbf{H}_{DMI} = \frac{2D}{\mu_0 M_S^2}((\nabla \cdot \mathbf{M})\hat{\mathbf{z}} - \nabla M_z) \tag{17}$$

Thus, the micromagnetic model contains the following interactions: Zeeman, uniaxial anisotropy with easy axis perpendicular to the track, direct exchange, DMI, and demagnetizing field. Additionally, surface roughness with 2 Å depth is introduced using an effective field model [52] as shown in Equation (18) – roughness is known to produce important effects on current-induced skyrmion motion, resulting in a variation of the skyrmion Hall angle with charge current density [53,54].

$$\mathbf{H}_{r,i} = -\left[\sum_j \mathbf{N}(i-j)G(i,j)\right]\mathbf{M}_i \qquad (18)$$

Here $G(i, j)$ is a function which models the surface roughness, with $\mathbf{N}$ being the usual demagnetizing tensor – for further details see Refs. [45,52]. A skyrmion collection is initialized in Figure 8(a) using an out-of-plane field of 2kA/m, following which a current density of $10^{11}$ A/m$^2$ is applied by setting a potential drop across the track, and the skyrmion motion is recorded under the influence of self-consistently computed spin torques. For a full list of material parameters for Co, Ir, and Pt see Ref. [45]. The computed skyrmion motion can then be compared with experimental results, as was done in Ref. [45] for a simpler Pt/Co bilayer with a smaller simulation space. With the multilayer track dimensions of Figure 8(a) such a comparison was not previously possible using single-GPU computation. However, this requires a separate publication and is left for a future work. Of interest here is the speedup factors which can be achieved by using such a multi-physics model. The demagnetizing interaction is now no longer the most expensive computational term, with the spin transport solver iteration dominating. Indeed, the speedup factors computed in Figure 8(b) as a function of number of spin transport solver cells (varied by changing the track in-plane dimensions) exceed those of the basic micromagnetic model, with efficiency factors reaching 0.96, 0.92, and 0.92 for 2, 3, and 4 GPUs respectively.

# VI. Conclusions

Here it was shown how micromagnetic and atomistic spin dynamics modelling can be accelerated using multiple GPUs, with the upgraded BORIS codebase available as open source [22]. Whilst the step from CPU to single-GPU computation provided a significant boost to performance, the step from single-GPU to multi-GPU computation unlocks the potential of modern hardware, allowing unprecedented performance both in terms of simulation size and speed. Key to this is a multi-GPU convolution algorithm, which has been introduced here to implement the long-range demagnetizing and dipole-dipole interactions efficiently. The algorithm splits the work equally between all available GPUs, with performance limited mostly by inter-GPU bandwidth. Even with a relatively cheap platform utilizing the PCIe 4.0 interface with 32 GB/s bandwidth, it was shown micromagnetic simulations can achieve speedup factors of 1.8, 2.5, and 3.1, for 2, 3, and 4 GPUs respectively, whilst atomistic spin dynamics can achieve speedup factors of 1.97, 2.9, and 3.9 for 2, 3, and 4 GPUs respectively. With 4 GPUs the maximum problem size is almost 4 times larger compared to the single-GPU case, and using 4 GPUs with 24 GB memory each, these were over 0.1 billion cells for 3D micromagnetic simulations, ~1 billion spins for atomistic spin dynamics, and ~2 billion spins for atomistic Monte Carlo simulations. It is important to stress the performance is only limited by the hardware used, with multi-GPU speedup efficiency largely limited by GPU latencies and inter-GPU bandwidth. The upcoming PCIe 6 platform with 128 GB/s bandwidth is predicted to allow over an order of magnitude speedup with 16 GPUs, compared to single-GPU computations. Modern GPU data servers allow 8, 16, or more GPUs interconnected using NVSwitch boards with 900 GB/s bandwidth for each GPU-to-GPU connection, which will increase in future generations. This potentially allows near-ideal speedup factors over 15 for 16 GPUs, as well as larger problem sizes.

## Data Availability

The data that support the findings of this study are available from the corresponding author upon reasonable request. The code discussed in this work is available in Ref. [22].

## Supplementary Material

The supplementary material contains additional information and technical details on micromagnetic interactions speedup efficiencies, effect of halved precision transfers, and use of additional NVLink bridges on performance.

# Appendix A

The multi-GPU convolution algorithm is presented in pseudo-code below. The implementation using C++17 is available as open source in Ref. [22].

---

**DATA**:

G : number of GPUs
$n_x, n_y, n_z$ : number of cells along x, y, and z dimensions respectively
$N_x = 2n_x$, $N_y = 2n_y$, $N_z = 2n_z$

Data partitioning $X_{ij}^{(g)}$ denotes memory stored on GPU $g$ (where $g$ is in the range 1, …, G), with $i$ denoting an equal partition along x dimension, and $j$ denoting an equal partition along y dimension ($i$ and $j$ are in the range 1, …, G).

Magnetization space **M** of dimensions $n_x \times n_y \times n_z$, stored on G GPUs as:
**M**$^{(1)}$ : $M_{11}^{(1)}$, …, $M_{1G}^{(1)}$
…
**M**$^{(G)}$ : $M_{G1}^{(G)}$, …, $M_{GG}^{(G)}$

Demagnetizing field space **H** of dimensions $n_x \times n_y \times n_z$, stored on G GPUs as:
**H**$^{(1)}$ : $H_{11}^{(1)}$, …, $H_{1G}^{(1)}$
…
**H**$^{(G)}$ : $H_{G1}^{(G)}$, …, $H_{GG}^{(G)}$

FFT input real space **In** of dimensions $N_x \times n_y \times n_z$, zero padded from $n_x + 1$ to $N_x$, stored on G GPUs as:
**In**$^{(1)}$ : $In_{11}^{(1)}$, …, $In_{G1}^{(1)}$
…
**In**$^{(G)}$ : $In_{1G}^{(G)}$, …, $In_{GG}^{(G)}$

FFT output real space **Out** of dimensions $N_x \times n_y \times n_z$, stored on G GPUs as:
**Out**$^{(1)}$ : $Out_{11}^{(1)}$, …, $Out_{G1}^{(1)}$
…
**Out**$^{(G)}$ : $Out_{1G}^{(G)}$, …, $Out_{GG}^{(G)}$

FFT complex space **S1** of dimensions $(n_x + 1) \times n_y \times n_z$, stored on G GPUs as:
**S1**$^{(1)}$ : $S1_{11}^{(1)}$, …, $S1_{G1}^{(1)}$
…
**S1**$^{(G)}$ : $S1_{1G}^{(G)}$, …, $S1_{GG}^{(G)}$

FFT complex space **S2** of dimensions $(n_x + 1) \times N_y \times N_z$, zero padded from $n_y + 1$ to $N_y$ and $n_z + 1$ to $N_z$, stored on G GPUs as:
**S2**$^{(1)}$ : $S2_{11}^{(1)}$, …, $S2_{1G}^{(1)}$
…
**S2**$^{(G)}$ : $S2_{G1}^{(G)}$, …, $S2_{GG}^{(G)}$

Convolution kernel **K** of dimensions $(n_x + 1) \times (n_y + 1) \times (n_z + 1)$, packed using y and z axis symmetries [24], and stored on G GPUs as: **K**$^{(1)}$, …, **K**$^{(G)}$

**PROCEDURES:**

**procedure** Transfer_M_to_In(**M**, **In**)
    **for** *i* in range 1, …, G **do**
        **for** *j* in range 1, …, G **do**
            **if** $i \neq j$ transfer $M_{ij}^{(i)}$ to $In_{ij}^{(j)}$, optionally halving precision before transfer
            **end if**
        **end for**
    **end for**
**end procedure**

**procedure** Transfer_Out_to_H(**Out**, **H**)
    **for** *i* in range 1, …, G **do**
        **for** *j* in range 1, …, G **do**
            **if** $i \neq j$ transfer $Out_{ij}^{(j)}$ to $H_{ij}^{(i)}$, optionally halving precision before transfer
            **end if**
        **end for**
    **end for**
**end procedure**

**procedure** Transfer_S1_to_S2(**S1**, **S2**)
    **for** *i* in range 1, …, G **do**
        **for** *j* in range 1, …, G **do**
            **if** $i \neq j$ transfer $S1_{ij}^{(j)}$ to $S2_{ij}^{(i)}$, optionally halving precision before transfer
            **end if**
        **end for**
    **end for**
**end procedure**

**procedure** Transfer_S2_to_S1(**S2**, **S1**)
    **for** *i* in range 1, …, G **do**
        **for** *j* in range 1, …, G **do**
            **if** $i \neq j$ transfer $S2_{ij}^{(i)}$ to $S1_{ij}^{(j)}$, optionally halving precision before transfer
            **end if**
        **end for**
    **end for**
**end procedure**

**procedure** ZeroPad(**S2**)
    **for** *i* in range 1, …, $N_G$ **do**
        Refresh zero padding of **S2**$^{(i)}$ from $n_y + 1$ to $N_y$ and $n_z + 1$ to $N_z$
    **end for**

**procedure** xFFT_R2C(**In**, **S1**)
    **for** *i* in range 1, …, $N_G$ **do**
        $n_y \times n_z$ real to complex FFTs with length $N_x$ of **In**$^{(i)}$ to **S1**$^{(i)}$
    **end for**
**end procedure**

**procedure** xFFT_C2R(**S1**, **Out**)
    **for** *i* in range 1, …, $N_G$ **do**
        $n_y \times n_z$ complex to real IFFTs with length $N_x$ of **S1**$^{(i)}$ to **Out**$^{(i)}$

```
            end for
    end procedure

    procedure yFFT(S2, direction)
            for i in range 1, …, N_G do
                    if direction is forward (n_x/G) × n_z FFTs with length N_y of S2^(i)
                    else (n_x/G) × n_z IFFTs with length N_y of S2^(i)
                    end if
            end for
    end procedure

    procedure zFFT(S2, direction)
            for i in range 1, …, N_G do
                    if direction is forward (n_x/G) × N_y FFTs with length N_z of S2^(i)
                    else (n_x/G) × N_y IFFTs with length N_z of S2^(i)
                    end if
            end for
    end procedure

    procedure KernelMultiplication(S2, K)
            for i in range 1, …, N_G do
                    Point-by-point complex multiplication of S2^(i) with K^(i), result stored in S2^(i)
            end for
    end procedure
```
**EXECUTION:**

Transfer_M_to_In(**M**, **In**)
xFFT_R2C(**In**, **S1**)
Transfer_S1_to_S2(**S1**, **S2**)
**if** in-place FFT ZeroPad(**S2**)
**endif**
yFFT(**S2**, forward)
zFFT(**S2**, forward)
KernelMultiplication(**S2**, **K**)
zFFT(**S2**, inverse)
yFFT(**S2**, inverse)
Transfer_S2_to_S1(**S2**, **S1**)
xFFT_C2R(**S1**, **Out**)
Transfer_Out_to_H(**Out**, **H**)

In the above algorithm the y and z FFTs may be executed in-place (same input and output memory spaces), which requires refreshing the zero padding. For improved performance, at the cost of extra memory, the FFTs may be executed out-of-place, which does not require refreshing the zero padding – this is the default option, with the code reverting to in-place FFTs when the GPU memory limit is reached. Additionally, it is beneficial to perform an xy transposition of the **S2** spaces to avoid use of strided y FFTs. This is not necessary for z FFTs since the z dimension is typically smaller, and due to use of pipelined convolution where the z FFTs, kernel multiplications and z IFFTs are done in the same routine (see Ref. [19]).

# Accelerating micromagnetic and atomistic simulations using multiple GPUs

## Supplementary Material

Serban Lepadatu[1]

[1]*Jeremiah Horrocks Institute for Mathematics, Physics and Astronomy, University of Central Lancashire, Preston PR1 2HE, U.K.*

## Speedup Efficiencies

Speedup efficiencies computed for the different interactions with iteration times shown in Figure 3 of the main text, are shown in Figure S1 for 2, 3, and 4 GPUs The computational speedup, S, is defined as the iteration time for single-GPU computation, divided by the iteration time for multi-GPU computation. The speedup efficiency is defined as Efficiency = (S – 1) / ($N_G$ – 1), where $N_G$ is the number of GPUs used. The maximum efficiency is 1 (ideal speedup), whilst negative efficiencies denote speedup values less than 1. For the single-site anisotropy interaction – Figure S1(a) – the maximum efficiency reached is 0.96 in all cases. GPU latencies affect the speedup efficiencies for smaller problem sizes, as the total latency increases by increasing the number of GPUs, however this becomes negligible for larger problem sizes above 10 million computational cells. The speedup efficiencies obtained for the local nearest-neighbour exchange interaction are shown in Figure S1(b). Again, the effect of latencies is observed at smaller problem sizes, with the maximum efficiencies now affected by inter-GPU memory transfers, resulting in maximum values of 0.93, 0.88, and 0.84 for 2, 3, and 4 GPUs respectively. Finally, the speedup efficiencies for the long-range demagnetizing interaction are shown in Figure S1(c). For a problem size with N computational cells, the total floating point numbers transferred over the bus is ~18N($N_G$-1)/$N_G$. Since the inter-GPU bandwidth is a significant bottleneck lower efficiencies are obtained in this case, with values of 0.54, 0.44, and 0.40 for 2, 3, and 4 GPUs respectively (corresponding to speedup factors of 1.54, 1.89, and 2.19 for 2, 3, and 4 GPUs respectively).

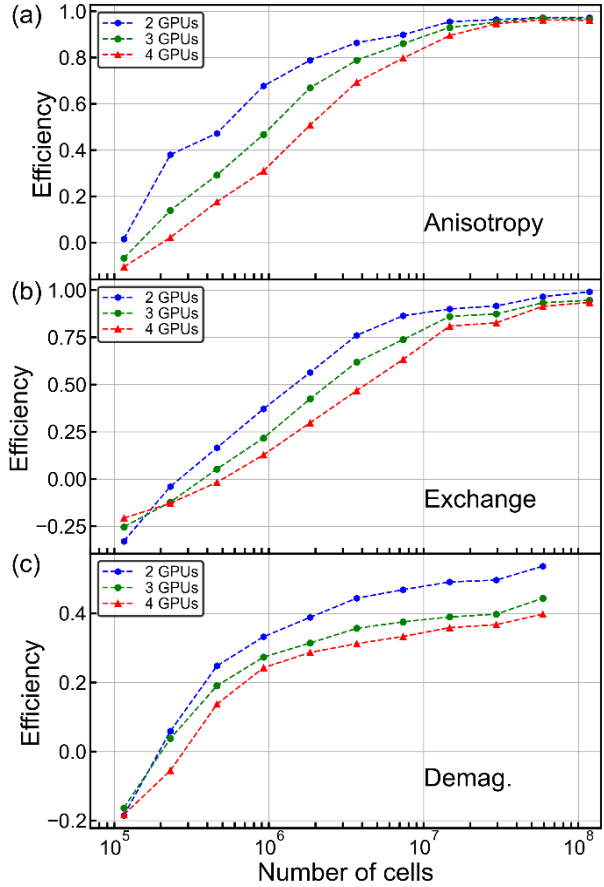

**Figure S1** – Speedup efficiencies for different interactions, including (a) single-site anisotropy, (b) local nearest-neighbour exchange interaction, and (c) long-range demagnetizing interaction.

## Effect of Half Precision Transfer on Performance

As discussed in the main text, for the multi-GPU convolution algorithm, data precision may be halved before transferring between GPUs. This helps when bandwidth is an important bottleneck. The effect of such mixed precision computation on the output error was already discussed in the main text. Here the effect on performance is shown, comparing the cases with and without halved precision transfers, plotted in Figure S2 for the demagnetizing interaction. As expected, iteration times increase by use of single precision transfers (single precision computations are used), rather than half precision – see panels (a) and (b) in Figure S2. However, for 2 GPUs the speedup factors are almost the same – compare panels (c) and (d) in Figure S2 – noting that conversion to and from half precision floating point numbers also introduces additional overhead. When used in a full micromagnetic model the increase in performance is negligible for 2 GPUs. For 3 and 4 GPUs the increase in performance is more significant, since in this case more data is being transferred overall between GPUs and bandwidth is a more important bottleneck. As discussed in the main text, in both cases the drop in efficiency between 3 and 4 GPUs is less than that between 2 and 3 GPUs – see panels (e) and (f) in Figure S2.

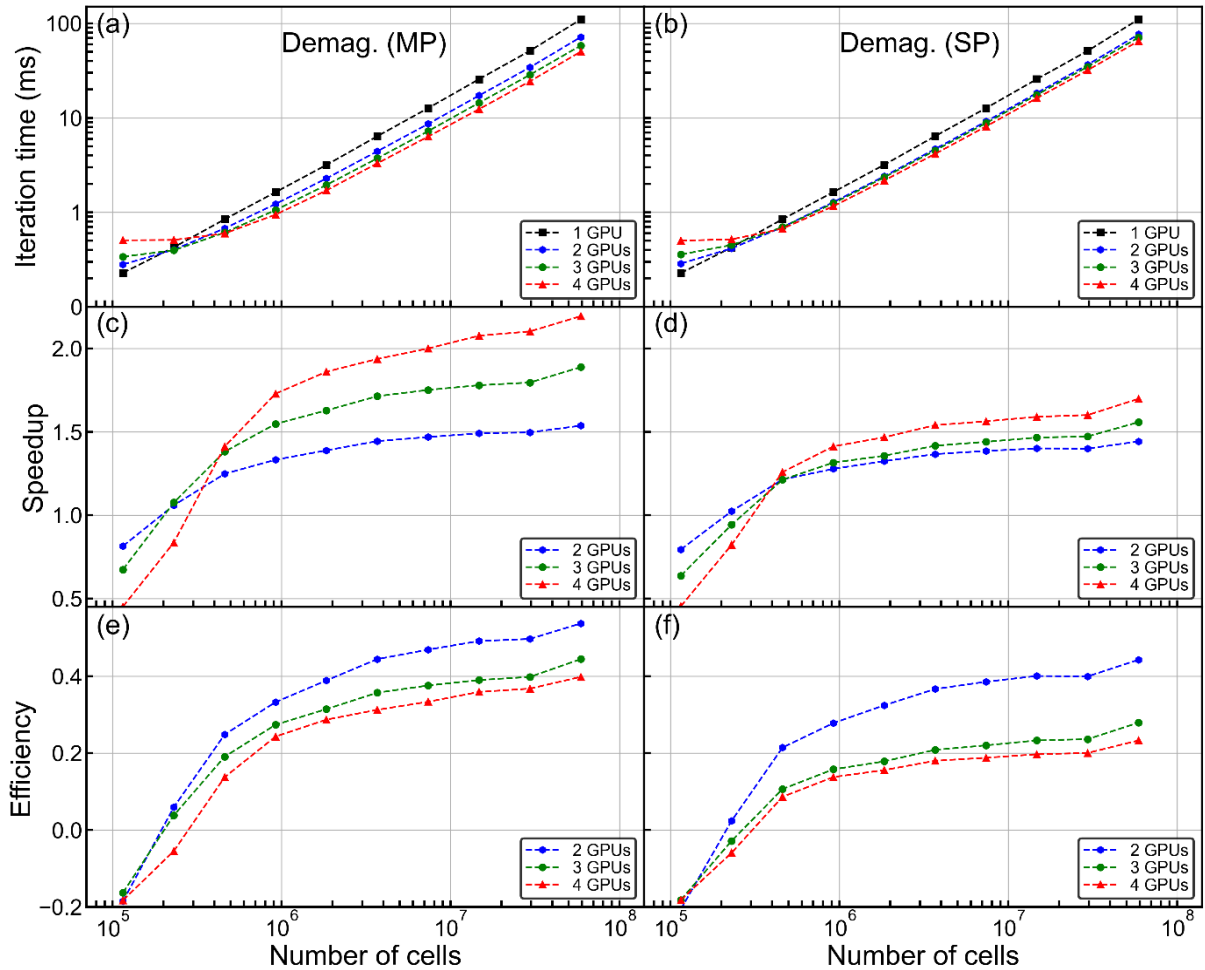

**Figure S2** – Performance comparison for the demagnetizing interaction for the case with mixed precision (MP) – single precision computation and half precision transfers – shown in the left-side panels, and fully single precision algorithm (SP), shown in the right-side panels. (a),(b) Iteration time, (c),(d) speedup, and (e),(f) speedup efficiency.

## Effect of NVLink Connections

As stated in the main text the computational platform used was a single workstation running Ubuntu 20.04, with 4 Nvidia A5000 24GB GPUs, connected via PCIe 4.0, each with an x16 slot (32 GB/s total bandwidth). The GPUs are also connected directly in pairs using NVLink bridges, which provides a separate higher 56.248 GB/s bandwidth. Use of additional

NVLink bridges not only increases bandwidth directly between the connected GPUs, but for the cases with 3 and 4 GPUs also reduces the total amount of data transferred over the PCIe bus. Thus, it is interesting to investigate the effect on performance of the improved connection topology.

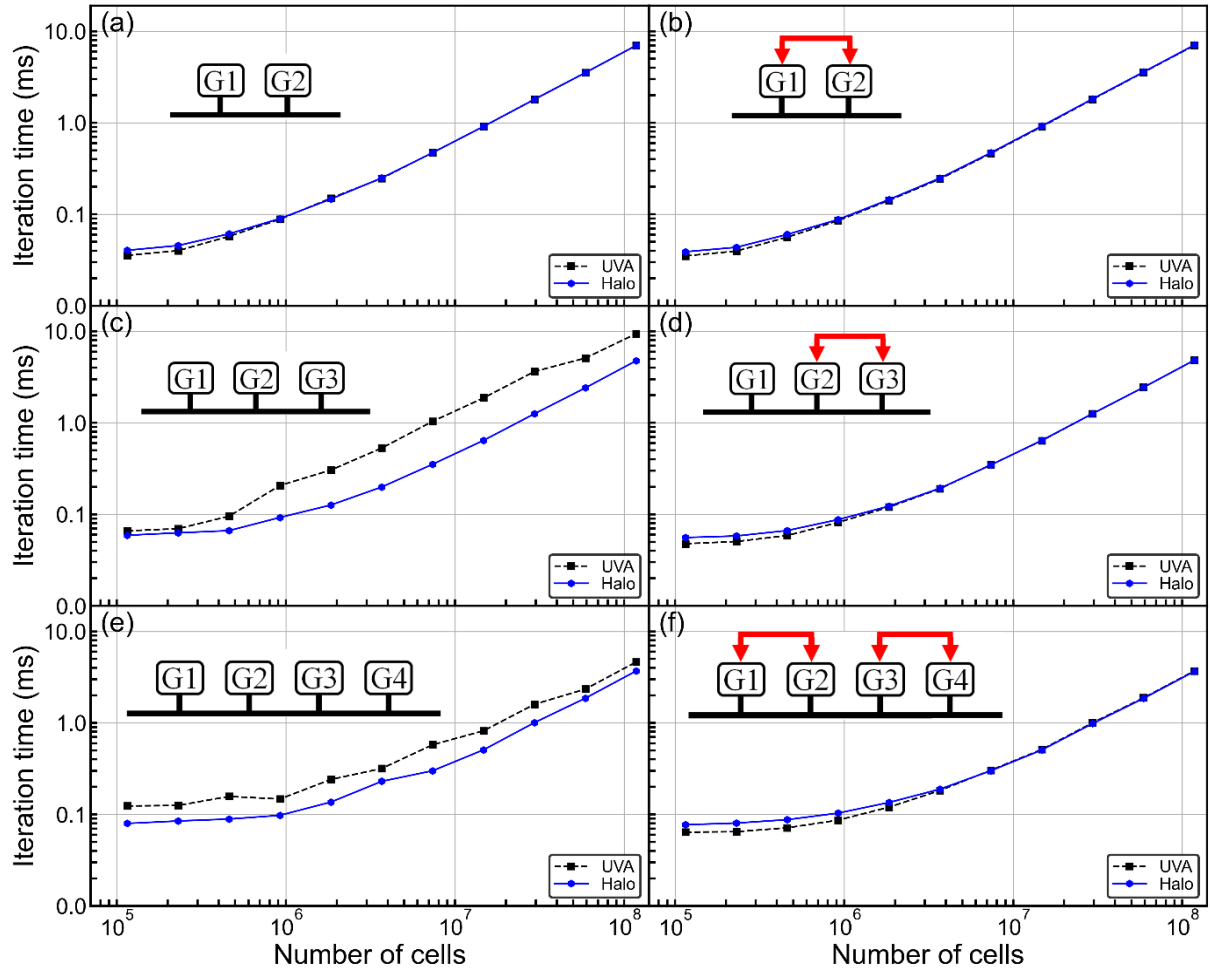

**Figure S3** – Iteration times for the exchange interaction, comparing the UVA and halo methods for different connection topologies. (a) 2 GPUs over PCIe bus, (b) 2 GPUs over NVLink bridge, (c) 3 GPUs over PCIe bus, (d) 3 GPUs with one NVLink bridge, (e) 4 GPUs over PCIe bus, and (f) 4 GPUs with 2 NVLink bridges. Solid black line denotes the PCIe 4.0 bus, and red lines with arrows denote NVLink bridges. G1, G2, G3, and G4 denote the different GPUs.

First, the effect on the exchange interaction iteration times is shown in Figure S3, where exchange interaction fields at GPU partition boundaries may be computed either using the UVA method or halo transfer method as discussed in the main text. The effect of additional bandwidth in this case is marginal since only a small fraction of the partitioned memory is being transferred between GPUs. Also as discussed in the main text the UVA method is better than the halo transfer method in most cases. There is however an exception to this, and that is when a computational routine on one GPU needs to access memory on 2 (or more) other GPUs using UVA over the PCIe bus. This is observed in Figure S3, panels (c) and (e), where the UVA method is significantly worse than the halo transfer method, also showing a slightly erratic performance. In all other cases, (a), (b), (d), and (f) in Figure S3, the UVA method outperforms the halo transfer method. Although not plotted, this is also the case for 4 GPUs where only the G2 and G3 GPUs are connected using an NVLink bridge. Further tests have been done using periodic boundary conditions, so that the first and last GPU also need to exchange data, and which support the same conclusion. As discussed in the main text, whilst memory transfers are done directly between GPUs over the PCIe bus, this is still a serial point-to-point connection, which requires CPU control to switch contexts. It is likely this additional overhead is the cause for poor performance when using UVA over PCIe.

Use of additional NVLink bridges does have a more important effect on the demagnetizing interaction computation performance, since here bandwidth is the largest bottleneck. This is shown in Figure S4 for 2, 3, and 4 GPUs. Use of additional NVLink bridges is observed to increase the measured speedup factors in all cases. The same comparison may be made for the micromagnetic problem of Figure 4 in the main text. The resulting speedup factors are shown in Figure S5. Use of NVLink bridges also results in larger speedup factors, however the increase in performance here is less significant, particularly for the case with 4 GPUs and 2 NVLink bridges, since the demagnetizing interaction share of the total computation time is reduced for this problem (see discussion in the main text).

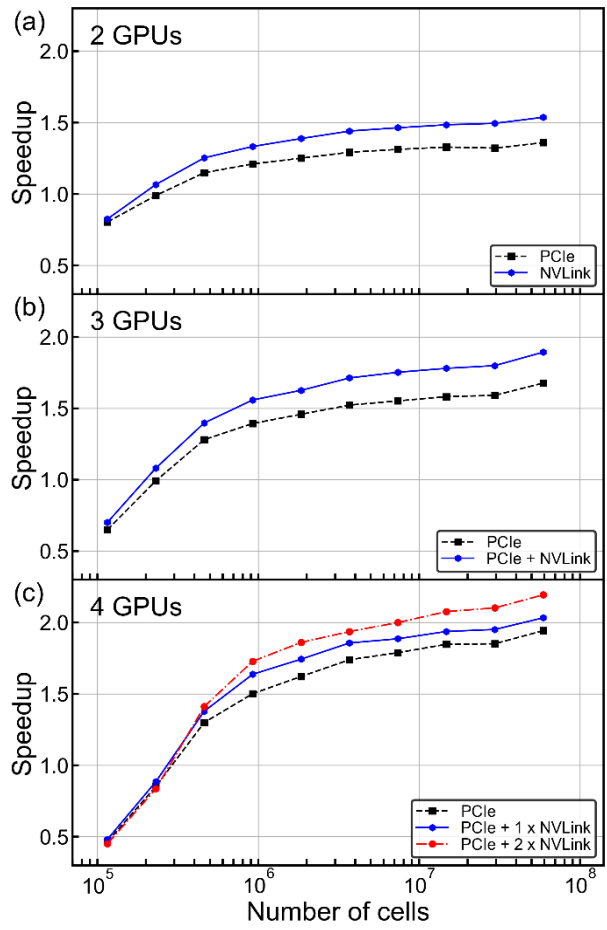

**Figure S4** – Demagnetizing interaction speedup factors for (a) 2, (b) 3, and (c) 4 GPUs, measured for different connection topologies, either all connected via PCIe bus, or with an additional NVLink bridge, or with 2 additional NVLink bridges.

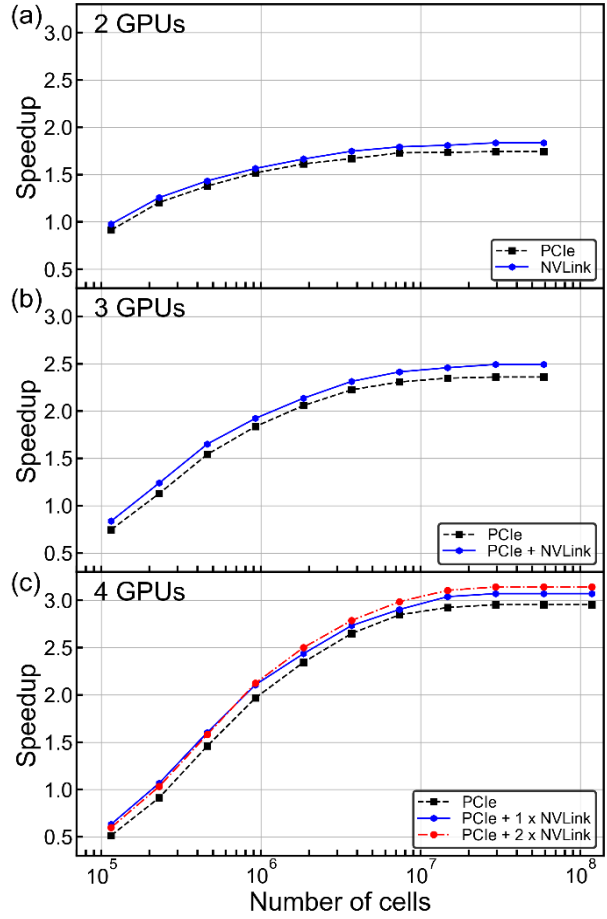

**Figure S5** – Speedup factors for the micromagnetic problem of Figure 4 in the main text, for (a) 2, (b) 3, and (c) 4 GPUs, measured for different connection topologies, either all connected via PCIe bus, or with an additional NVLink bridge, or with 2 additional NVLink bridges.

## Parallelization Speedup Model

The number of floating point values transferred each iteration is given by $18N(N_G - 1)/N_G$, where N is the number of computational cells and $N_G$ the number of GPUs. With a data transfer rate *R* (e.g. 56.248 GB/s for NVLink), and number of bytes per floating point value *B* (*B* = 2, 4 for half and single precision respectively), the time spent each iteration transferring data is given by:

$$t_R(N_G) = \frac{18N(N_G-1)}{N_G}\frac{B}{R}$$

Then, if $t_I$ is the time spent evaluating interactions which can be parallelized efficiently (Zeeman, anisotropy, exchange, LLG evaluation), and $t_D$ is the time spent evaluating the demagnetizing interaction, the total time for each iteration is:

$$t_T(N_G) = \frac{t_I + t_D(r_D + r_{Dextrap})}{\eta N_G} + t_R r_D$$

Here $\eta$ is an efficiency factor (<1.0), and $r_D$, $r_{Dextrap}$ are demagnetizing evaluation time reduction factors introduced when using the polynomial extrapolation method. Thus for RKF56 $r_D = 1/8$ since the method contains 8 sub-steps per iteration, and $r_{Dextrap} = 0.1$ is an additional factor introduced due to polynomial extrapolation computation; for RK4 we have $r_D = 1/4$ and $r_{Dextrap} = 0.05$. The efficiency $\eta$ can be estimated by fitting the measured interaction times for the results up to 4 GPUs. Thus for a problem with N = 58,982,400 computational cells, measured interaction times are $t_I$ = 26.8 ms, $t_D$ = 122.3 ms, and $\eta$ = 0.98 is obtained. Results are plotted in Figure S6.

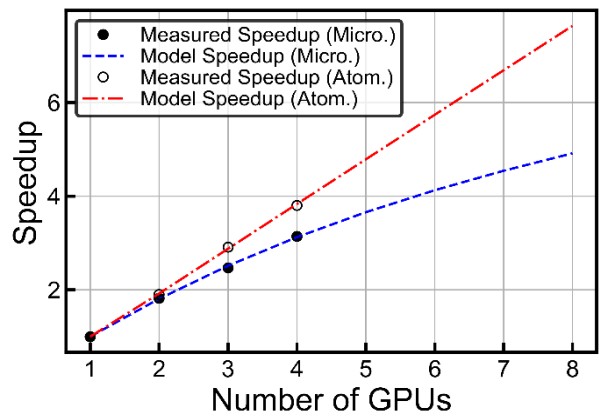

**Figure S6** – Speedup factors as a function of number of GPUs predicted by the parallelization model for the micromagnetic problem of Figure 5 in the main text (dashed line), and atomistic problem of Figure 6 in the main text (dash-dot line). The solid and open symbols are measured speedup factors for the micromagnetic and atomistic problems respectively.

For the micromagnetic problem with RKF56 evaluation the model predicts a speedup factor close to 5 is possible for 8 GPUs, although it is likely this can be achieved only for larger problem sizes. In this case, due to the restrictive bandwidth, parallelization efficiency is low and gains obtained with more than 4 GPUs are limited. For atomistic spin dynamics, where spin-averaged cells are used to evaluate the demagnetizing interaction, this can be further improved. The same formula for $t_T$ applies, however $t_D$ is now significantly reduced; e.g. if $N_S$ is the number of atomistic spins in each spin-averaged macrocell ($N_S = 64$ for the atomistic problem in Figure 6 of the main text), then $t_D$ is reduced, and more importantly $t_R$ is also reduced, given now by:

$$t_R(N_G) = \frac{18N(N_G - 1)}{N_S N_G} \frac{B}{R}$$

For the same problem size we now have $t_D = 7.2$ ms, $t_I = 161.7$ ms (this now includes additional contributions due to 2-temperature heat solver iteration and pseudo-random number generation for the sLLG equation), and $\eta = 0.96$ is obtained. Since $t_I$ is parallelized efficiently, as seen in Figure S6, and $t_R$ is much smaller, then the model predicts a near ideal speedup for 8 GPUs (speedup factor of 7.5), although again this likely requires larger problem sizes.

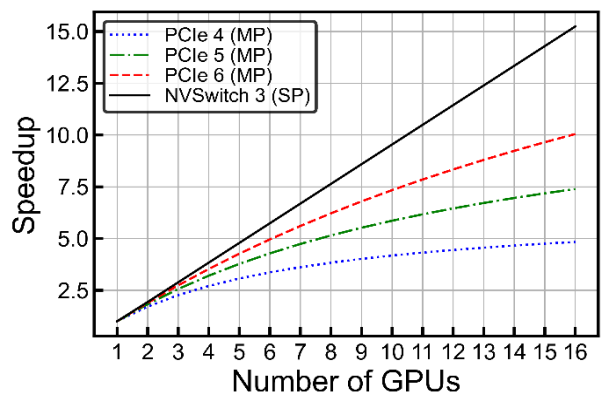

**Figure S7** – Speedup factors as a function of number of GPUs predicted by the parallelization model for different platforms. For PCIe 4, 5, and 6 the micromagnetic problem of Figure 5 in the main text is used (mixed precision with RKF56 evaluation and polynomial extrapolation for the demagnetizing field), whilst for the NVSwitch 3 platform fully single precision convolution algorithm and no polynomial extrapolation are used (B = 4, $r_D = 1$, $r_{Dextrap} = 0$) – two connected NVSwitch boards are assumed, which accommodate up to 16 GPUs.

It is instructive to apply the developed parallelization model to predict speedup factors for different bandwidths. First we consider the PCIe platform, namely versions 4, 5, and 6, which provide bandwidths of 32 GB/s, 64 GB/s, and 128 GB/s respectively (PCIe 6 is due to be released in 2024, and it should be noted PCIe 7 is projected to provide 256 GB/s). Results are shown in Figure S7, again for the micromagnetic problem with RKF56 evaluation and polynomial extrapolation. As can be seen, the PCIe 4 platform does not provide enough bandwidth to make computations with over 3 or 4 GPUs efficient, although for 2 GPUs the obtained speedup factor of >1.7 is significant. As the bandwidth is increased, larger problem sizes with increasing number of GPUs can be simulated more efficiently, and it is predicted for PCIe 6 a speedup factor of ~10 can be obtained with 16 GPUs.

Finally, the NVSwitch platform provides for full interconnections between GPUs (i.e. each GPU is connected to all other GPUs directly), which means $t_R$ can also be parallelized – for PCIe, which is a serial connection, only one transfer between 2 GPUs can be made at any one time, however for NVSwitch asynchronous data transfers between GPUs can be configured. Now $t_R$ is the required time for 1 GPU to transfer data to all other GPUs, which is given by:

$$t_R(N_G) = \frac{18N(N_G-1)}{N_G^2}\frac{B}{R}$$

The NVSwitch 3$^{rd}$ generation platform allows for 8 GPUs to be interconnected, each with 900 GB/s bandwidth (7.2 TB/s aggregate bandwidth), however multiple NVSwitch boards can also be interconnected. Using the fully single precision convolution algorithm ($B = 4$), and no polynomial extrapolation for the demagnetizing field evaluation ($r_D = 1$ and $r_{Dextrap} = 0$), the results in Figure S7 predict a nearly ideal speedup for up to 16 GPUs (~15 speedup factor for 16 GPUs), although again, the usefulness of a large number of GPUs is likely realised only for large problems.

## Perpendicular Magnetic Tunnel Junction Problem

For the problem of Figure 3 in the main text, the resistance in the tunnel barrier is calculated using:

$$R(\theta) = \frac{R_0}{1+\frac{R_{ap}-R_p}{R_{ap}+R_p}\cos\theta}$$

Here $R_0 = 2R_{ap}R_p/(R_{ap}+R_p)$ is the resistance for perpendicular orientation of ferromagnetic layers' magnetizations, with $R_{ap}$ and $R_p$ being resistances for antiparallel and parallel orientations respectively. The tunnelling magnetoresistance (TMR) ratio is defined as $TMR = (R_{ap} - R_p)/R_p$, and the resistance-area product is defined as $RA = R_p A$, where $A$ is the cross-sectional area of the MTJ.

Exchange bias coupling between the permanent magnetic layer (PL) and the antiferromagnetic (AFM) layer is included using:

$$\mathbf{H}_{ij} = \frac{J_A}{\mu_0 M_S h_z} \mathbf{m}_{j,A}$$

Here cell $i$ in PL is coupled to cell $j$ in AFM, such that $\mathbf{m}_{jA}$ is the magnetization direction of sub-lattice A in AFM, $h_z$ is the z cellsize of PL, and the coupling constant is $J_A = 1$ mJ/m$^2$.

Surface exchange coupling between the free layer (FL) and PL is included similarly using:

$$\mathbf{H}_{ij} = \frac{J_1}{\mu_0 M_S h_z} \mathbf{m}_j$$

Here the coupling constant is $J_1 = 0.1$ mJ/m$^2$.